# Phonon Magic Angle in Two-Dimensional Puckered Homostructures

*Yufeng Zhang[†, §], Meng An[†, ‡, §], Dongxing Song[†], Haidong Wang[†], Weigang Ma[†],*

*Xing Zhang[\*, †]*

[†] Key Laboratory for Thermal Science and Power Engineering of Ministry of

Education, Department of Engineering Mechanics, Tsinghua University, Beijing

100084, China

[‡] College of Mechanical and Electrical Engineering, Shaanxi University of Science

and Technology, Xi'an 710021, China

[§] Y. Z. and M. A. contributed equally to this work

[\*] Corresponding author. E-mail: x-zhang@mail.tsinghua.edu.cn

ABSTRACT

The emergence of twistronics provides an unprecedented platform to modulate the band structure, resulting in exotic electronic phenomena ranging from ferromagnetism to superconductivity. However, such concept on phonon engineering is still lacking. Here, we extend the 'twistnonics' to 2D puckered materials with a 'phonon magic angle' discovered by molecular dynamics simulation. The phonon magic angle, with the TP-1 and TP-2 direction overlapped, remains a high level or even enhances phonon transport capability due to van der Waals confinement. This novel phenomenon originates from the confined vdW interaction and ordered atomic vibration caused by the perfect lattice arrangement that the atoms of the top layer can be stuck to the spaces of the bottom layer. Moreover, it is found that both the in-plane and out-of-plane thermal transport properties can be effectively regulated by applying the twist. Through the phononic and electronic analysis, the deterioration of phonon transport capability for other twist angles are attributed to the suppression of acoustic phonon modes, reduction of phonon lifetimes and mismatched lattice vibration between layers. Our findings shed light on the twistnonics of low-dimensional asymmetrical materials and can be further extended to electronic and photonic devices.

Recently, the rapid development of twistronics enables the breakthrough of vast electronic behavior in two-dimensional (2D) materials.[1-5] It is known that van der Waals heterostructures formed by vertical or lateral stacking offer new characteristics and broaden the material phase space in 2D world.[6-10] Interlayer twist, however, is another approach to forming 2D heterostructures or homostructures through self-assembly. By varying the relative angle between layers, a variety of different high-symmetry configurations can simultaneously appear, which are so-called Moiré patterns.[11-13] This phenomenon induced by lattice mismatch acts as a superlattice modulation and provides an unprecedented platform to modulate physical properties, including superconductivity, ferromagnetism, Hofstadter butterfly and topological valley transport.[14-17] As a prominent example, twisted bilayer graphene (TBG) has been widely studied with groundbreaking results ranging from visible absorption to superconductivity.[18-21] In addition, other twisted 2D materials such as transition metal dichalcogenides (TMDs) also show angle dependent electronic and optical properties.[22,23] With the development of material science, the preparation technique of twisted heterostructures has become mature and the twist accuracy can be precisely controlled within 0.1°, which paves the way for future twistronic researches.[24-26]

Although graphene has excellent electronic, mechanical, optical and thermal properties, the lack of band gap blocks its application in semiconductor field. In searching for the substitutes that beyond graphene, 2D puckered materials emerged as rising stars due to their tunable band gap and extraordinary properties.[27, 28] This new class of 2D materials have regular fluctuated structure in armchair direction and mainly includes group V elemental films (e.g. black phosphorus) and group IV monochalcogenides (e.g. SnSe, GeS). Thanks to the corrugated configuration, 2D puckered materials exhibit strong in-plane anisotropy and are regarded as potential candidates in field-effect transistors (FETs), photodetectors, piezoelectric and thermoelectric fields,[29-32] all of which are inseparable from heat conduction. As the rules of heat transfer in semiconductor materials are dominated by phonons, the effective modulation of phonons is a clear and ever-growing need in nanoscale thermal science. The van der Waals (vdW) interaction that exists universally in 2D homo or heterostructures is hailed as a highly attractive platform for phonon regulation because of its tunable flexibility to individual components, phonon coupling strength and vdW interlayer distance. For example, the phonon-phonon scatterings caused by the vdW interaction can block the phonon transport in nanostructures.[33, 34] On the other hand, the thermal conduction can also be enhanced by vdW confinement in polyethylene single chain,[35] h-BN nanoribbon[36] or quasi-1D van der Waals crystal $Ta_2Pd_3Se_8$ nanowires.[37]

Twistnonics, analogous to twistronics for phononic engineering, offers a new opportunity to successively and effectively manipulate the vdW interaction in 2D nano-phononic devices. The interlayer twist essentially changes the interaction of atoms and potential energy between layers, thus leading to the modulation of band structures and transport properties.[38-40] However, as the study of twistronics has been promoted rapidly, the effect on the twistnonics has received the least attention and least understood [41-44]. Among them, TBG attracted the most attention both experimentally and theoretically, which shows a lower thermal conductivity than Bernal stacking due to enhanced phonon Umklapp and normal scattering.[41, 42] The angle dependent thermal conductivity of TBG was found with a local maximum value during the twist angle of 30°.[43] However, the same twist angle shows a minimum thermal conductivity of h-BN supported monolayer graphene.[44] Except for the graphene, twisted $MoS_2$/$MoSe_2$ heterostructures were also investigated but found to have little impact on the thermal conductivity.[45] Recent progress in twistnonics indicated that low-frequency modes are extremely sensitive to twisting in the twisted TMD bilayer.[46] This angle dependent characteristic of phonon can be used as a probe to determine the twist angle. Although several studies have been conducted on the graphene and TMDs, the exploration of phononic modulation in twisted 2D puckered materials is still lacking, which opens the door to the twistnonics of 2D anisotropic world.

In this work, we extend the twsitnonics to 2D puckered homostructures by molecular dynamics (MD) simulation. We first define the magic angle of 2D puckered materials and verify its thermodynamic stability though the radial distribution function. Based on morphological analysis, twisted bilayer black phosphorous (TBBP) is taken as an example to study the phononic properties. Then the thermal conductivity in both orthogonal directions and the interfacial thermal resistance with respect to various twist angles are systematically investigated. Phonon transmission coefficient and spectral energy density are employed to analyze the contribution of phonons and phonon lifetimes of different frequencies to in-plane thermal transport. Moreover, electronic density of various twist angles are calculated to reveal the heat conduction performance at the interface by the behavior of electrons.

RESULTS AND DISCUSSION

Morphological stability analysis

It is well known that AB stacking is the most common and stable stacking manner for multi-layer 2D materials, regardless of whether the surface is flat (e.g. graphene, BN) or puckered (e.g. BP, SnSe). Among them, the 2D puckered materials with AB stacking is formed by conducting a translation transformation of half of the lattice parameters from AA stacking in both the armchair and zigzag directions, so that the atoms in top layer can get stuck to the interspace of bottom layer. Apart from the widely studied zigzag and armchair directions, there are two other principle directions that along the diagonal of the lattice, which are called third principle (TP-1 and TP-2) directions here (see Figure 1(a) and (b)).[47] If the top layer is twisted by a certain angle $α$ (or 180-$α$), its TP-1 (or TP-2) direction can be coincident with the TP-2 (or TP-1) direction of the bottom layer. In this case, it is interesting to find that the atoms of the top layer can also be completely confined to the space between atoms in the bottom layer due to the same interatomic spacing in the TP-1 and TP-2 direction, which is similar to the AB stacking. In this situation, we define this special angle, $\alpha$ ($0 \leq \alpha \leq 90$), as 'magic angle' in 2D puckered structures.

To describe the structure of magic angle, the morphology is qualified to show the atomic vibrations through radial distribution function (RDF).[48] In simulation studies, RDF usually serves as an effective indicator of the nature of the phase (e.g. solid, liquid and gas) and the crystal lattice structures (e.g. fcc, bcc and hcp). Twisting can change the lattice structure, the arrangement of atoms in each layer, however, remains unchanged. So the conventional RDF defined as the probability of finding an atom at a certain distance from another tagged atom will not characterize such unique difference. Therefore, to clearly analyze the morphological stability of twisted 2D puckered structures, the modified RDF is defined as

$$\rho g(r) = \frac{2}{N} \left\langle \sum_{i}^{N_T} \sum_{j}^{N_U} \delta\left[\mathbf{r} - \mathbf{r}_{ij}\right] \right\rangle \quad (1)$$

where $g(r)$ is the RDF, $\rho$ is the number density, $\mathbf{r}_{ij}$ is the in-plane vector between centers of atom $i$ in twisted layer and atom $j$ in untwisted layer, $N_T$ and $N_U$ are the number of atoms of twisted layer and untwisted layer, respectively, and the angular brackets represent a time ensemble average. The main modification is that $\mathbf{r}_{ij}$ does not include the vectors of atoms in the same layer but only considers that of atoms in different layers, so the relative arrangement of atoms between layers can be better characterized. To be typically, we show the RDF results for BP and SnSe in Figure 1(c) and 1(d), respectively, with four angles that are original AB stacking, magic angle ($\theta$=70.5° for BP and $\theta$=86.4° for SnSe) and two other arbitrary angles. It can be found that the $g(r)$ of AB stacking is zero when $r$ is close to the original point and has periodic sharp peaks for both BP and SnSe due to the orderly stacked structure. Interestingly, the magic angle also owns a zero part near the origin, which indicates that there is no overlap between atoms in the top and bottom layers during the process of thermal motion. In addition, the regular peaks can also be observed in magic angle but with a wider FWHM (half width at half maximum). The peaks existed in RDF represent the distance between each neighbours and the lattice vibration around the equilibrium positions, while the wider peak of magic angle is mainly attributed to the stronger vibration of atoms due to the fact that the interlayer friction resistance is smaller that of the AB stacking, yet much larger than other twist angles.[49] The atomic structure in pairs from the in-plane perspective may leads to the superposition of two Gaussian peaks, thus further widening the peaks. Compared to the AB stacking and magic angle, $g(r)$ of arbitrary angles shows a non-zero beginning and they quickly approach to the normal bulk density with $g(r) = 1$. This comparison can be clearly seen in the lattice structures below the figures, where the atoms are highly ordered in AB stacking and magic angle. To be brief, the RDF analysis reflects the ordered lattice structure and high degree of matching of atoms in top and bottom layers when bilayer puckered material is twisted by magic angle.

To further verify the thermodynamics stability of magic angle, we conducted a MD simulation which imitates the *in situ* cleavage experiment.[50] Initially, a small square piece with length of 2.1 nm is placed above the middle of a large square piece with length of 10 nm, and the initial twist angle is set to be an arbitrary value between 0° and 90°. After minimization, the whole system under periodic boundary condition was relaxed in the isothermal-isobaric (NPT) ensemble to reach a stable thermodynamics state. As a typical example, we recorded the comprehensive thermodynamics process for BP to reach the stabilized stacking of magic angle in the Supporting Information (magic-BP.avi). It is clearly seen that the small square piece shifts with respect to the substrate piece to search for a stable site. During searching process, variable Moiré patterns can be observed induced by thermodynamics twist. An angle of 70.5° is stabilized after full relaxation, which verifies the stability of the stacking of magic angle and it can be directly obtained by *in situ* cleavage for 2D puckered homostructures.[49] For comparison, the thermodynamics process for AB stacking is also given in the Supporting Information (AB-BP.avi). We found that the final stable stacking is determined by both the initial angle and the velocity vector of atoms. In a word, besides the well-known AB stacking, the magic angle is another stable stacking manner, which is applicable for all the twisted 2D puckered materials.

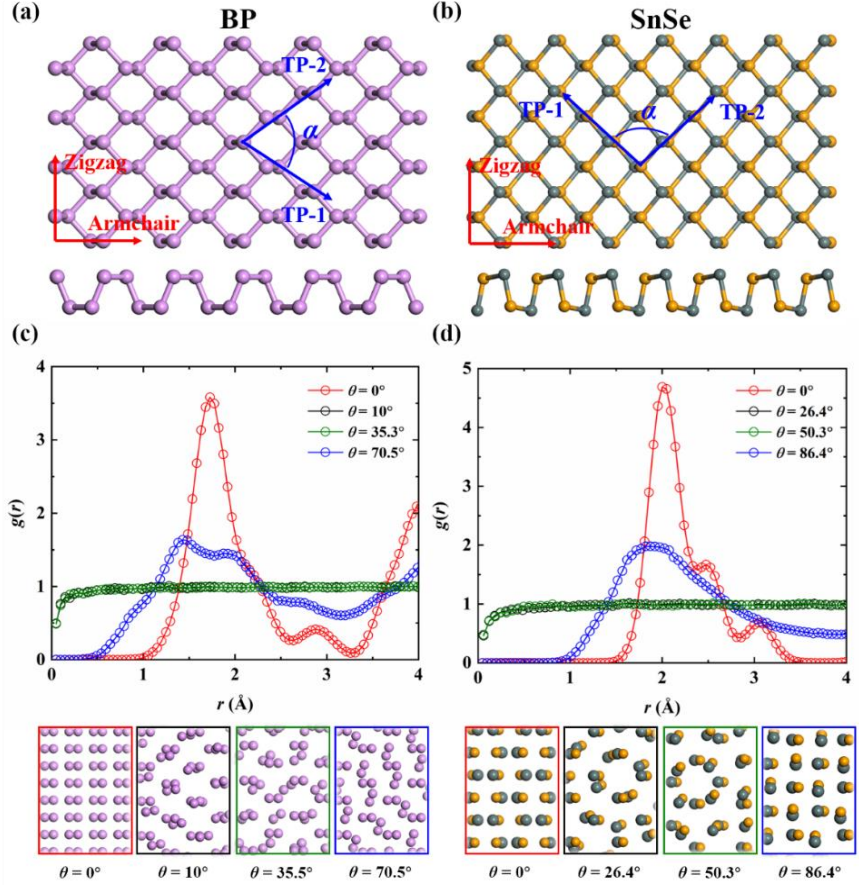

**Figure 1.** Morphological structures of (a) BP and (b) SnSe, respectively, where the red arrows represent the conventional zigzag and armchair direction, and the blue arrows represent the newly defined TP-1 and TP-2 direction. (c) RDF of twisted BP with $\theta = 0°$, $\theta = 10°$, $\theta = 35.5°$ and $\theta = 70.5°$. (d) RDF of twisted SnSe with $\theta = 0°$, $\theta = 26.4°$, $\theta = 50.3°$ and $\theta = 86.4°$. The rectangular patterns below each figures are the atom arrangements of four different angles for clear comparison. The frames with color of red, black, green and blue corresponds to AB stacking, the first arbitrary angle, the second arbitrary angle and the magic angle, respectively, and is inconsistent to the color of curves.

In-plane thermal conductivity

We calculated the in-plane thermal conductivity of TBBP in both *x* and *y* directions due to its anisotropic characteristics. Only the thermal conductivity along *y*-axis is discussed here to avoid duplication and for its vital significance in thermoelectric application (See Figure S1 for thermal conductivity along *x*-axis).[32, 51] As shown in Figure 2(a), the armchair direction of the top layer (untwisted layer) is kept in consistent with the *y*-axis and the heat flux direction (*y* direction), while that of the bottom layer (twisted layer) can be varied in TBBP. The thermal conductivity of bilayer, twisted layer and untwisted layer are shown in Figure 2(b), (c) and (d), respectively. It can be found that the overall thermal conductivity of bilayer, which is decided by both layers, exhibits an upward trend with twist angle varied from 0° to 90°. To discuss the thermal conductivity of each layer separately, the twisted layer can be regarded as a monolayer BP supported on a constant substrate with different twist angles. The highly anisotropic thermal conductivity of BP is mainly attributed to significantly different acoustic phonon bandwidths and group velocities between Γ-X (zigzag) and Γ-Y (armchair) directions shown in Figure S2. Such the direction-dependent group velocities and anisotropic phonon dispersion relation have also been demonstrated in previous studies.[52] Therefore, the thermal conductivity of twisted layer shows an increasing trend since that the direction along the *y*-axis gradually transformed from the armchair with the lowest conductivity to the zigzag with the highest value, which agrees with the cosine function fitting of the chirality-dependent thermal conductivity of monolayer

BP.[53] Surprisingly, a singularity was found counterintuitively at $\theta = 70.5°$, so as the thermal conductivity of bilayer, which exactly corresponds to the magic angle. Furthermore, the thermal conductivity of untwisted layer shows a complicated trend since it can be treated as a monolayer BP with fixed crystal orientation supported on a twisted substrate. Except for $\theta = 0°$, the thermal conductivity also shows a cosine function of twist angle and owns a maximum value at magic angle that even higher than AB stacking. Therefore, considering the superior thermal properties of magic angle, we further refer to it as 'phonon magic angle'. As the morphology we discussed before, the characteristic peak in RDF stands for the highly ordered structure of magic angle caused by the vdW confinement. The confined structure restricts the lattice vibration and may further result in the superb behavior of phonons.

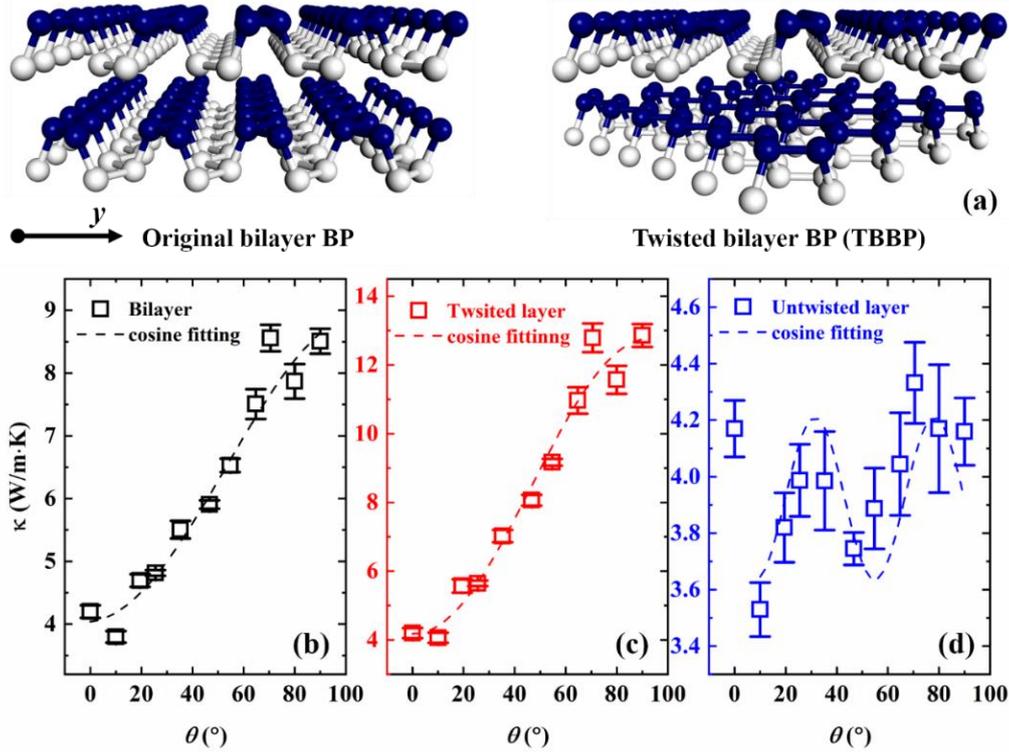

**Figure 2.** (a) Concept diagram for bilayer BP of AB stacking and with twist angle. The top layer (untwisted layer) is always fixed with zigzag and armchair direction corresponding to the *x*-axis and *y*-axis, respectively. The bottom layer (twisted layer) is applied twist to a certain angle. The different colors of P atoms are for clearly distinguishing the morphology. Below are the thermal conductivity results for (b) bilayer, (c) twisted layer and (d) untwisted layer of various twist angles from 0° to 90° along the *y* direction with black, red and blue staining, respectively. The dashed lines are fitted by cosine function for visual guide.

To understand the effect of twist on thermal conductivity of the untwisted layer and the physical mechanism behind the phonon magic angle, we calculate the spectral phonon transmission coefficient including full orders of phonon-phonon interactions in NEMD simulations for specific angles that are 0°, 10° and 70.5° using the method developed by Sääskilahti et al.[54, 55] (see Part S3 for calculation details). In Figure 3(a), the calculation results demonstrate that for three specific angles phonons have the same range of frequency from 0 THz to 13.5 THz with a significant band gap around 9 THz that reflects the semiconductor characteristics of BP. In conventional thermal regulation approaches such as defect and doping engineering, the extremely low frequency phonons near the Brillouin zone center exhibits the similar behavior or suffer a reduction compared with the original structure.[56-58] However, a small peak is clearly observed in TBBP close to 0 THz for both $\theta = 10°$ and $\theta = 70.5°$, which means that the low-frequency modes can be stimulated by applying the twist. This intriguing phenomenon has also been reported in twisted TMD bilayer where low-frequency shear modes (SMs) and layer breathing modes are extremely sensitive to twisting and can be used to infer the twist angle.[46] Beyond the Brillouin zone center, the transmission coefficient of $\theta = 10°$ is significantly lower than that of $\theta = 0°$ in low frequency mainly ranging from 1 THz to 5 THz, which indicates the suppression of acoustic modes and part of the optical modes. It is noteworthy that such reduction in thermal conductivity has also been widely observed in TBG due to the enhanced phonon Umklapp and normal scattering within the low-energy range.[41-44] In contrast, it is interesting to find that the transmission coefficient of $\theta = 70.5°$ (magic angle) maintain the same level as that of $\theta = 0°$ (AB stacking) in the entire spectrum range, which can explain for the considerable value of thermal conductivity. This phenomenon can be attributed to the relatively small commensurate primitive lattice constant of magic angle, which maintains the Brillouin zone size and avoids the emergence of folded acoustic phonon branches.[41, 42] Besides, looking back to the morphology of magic angle, it can be found that the lattice vibration is confined in the potential barriers by ordered vdW interaction, thus leading to reduction of atomic thermal displacement magnitudes and less phonon scattering.[35] Therefore, the highly symmetrical and confined structure maintains the size of the Brillouin zone and reduces the scattering, thereby further realizing the prodigious performance of phonons, as the so-called phonon magic angle.

On the basis of spectral phonon transmission coefficient, we will discuss the contribution of phonons of different frequencies to thermal conductivity. Figure 3(b) is obtained by integrating the transmission coefficient. Consistent with previous analysis, the reduction in thermal conductivity is mainly concentrated in 1-5 THz. Moreover, the difference between $\theta = 0°$ and $\theta = 70.5°$ is magnified. It can be clearly seen that the slightly higher thermal conductivity of $\theta = 70.5°$ is due to the more contribution of optical phonons (5-6 THz), especially the $B_{1g}$ mode.[59] In conclusion, the suppression of low frequency acoustic phonons due to enhanced phonon scattering leads to the attenuation of the thermal conductivity of untwisted layer in TBBP. In contrast, the nano-confinement existed in phonon magic angle caused by the ordered vdW interaction reduces the phonon scattering, thus resulting in a slight enhance in optical modes and even higher thermal conductivity than conventional AB stacking.

To gain further understanding of the discrepancy of phonon lifetimes and dispersion relation, the normalized spectral energy density (SED) that involves the full anharmonicity of lattice interactions is calculated (see Part S4 for calculation details). The SED for $\theta = 0°$, $\theta = 10°$ and $\theta = 70.5°$ in the full-frequency range is given in Figure S4. Keeping in view that the dominant thermal energy carriers in BP are low frequency phonons according to the aforementioned analysis, an enlarged SED contour map in the frequency range below 5 THz is presented in Figure 3(c). As denoted by white solid ellipses, phonon brunches of $\theta = 10°$ show significant broadening and become more indistinct near 2 THz comparing to AB stacking and magic angle. This broadening effect implies the reduction of phonon lifetimes due to the introduced phonon scattering and supports the suppression of the acoustic mode at low frequencies reflected in the transmission coefficient analysis. Moreover, the lower group velocity indicated by the slope of the acoustic phonon brunches at the Gamma point (shown as solid straight lines) also leads to the low thermal conductivity. Additionally, new phonon branches are found to appear near the Brillouin zone center in non-zero twisted structures as denoted by white dashed ellipses, which agrees with the peaks close to 0 THz in Figure 3(a). The especially distinct branches for magic angle suggest the larger phonon lifetimes and can be responsible for its superior thermal conduction. Therefore, we have further discussed the phonon characteristics from the aspect of lifetimes and found that the result is consistent with the previous analysis and phonon magic angle owns relatively long lifetimes at low frequency.

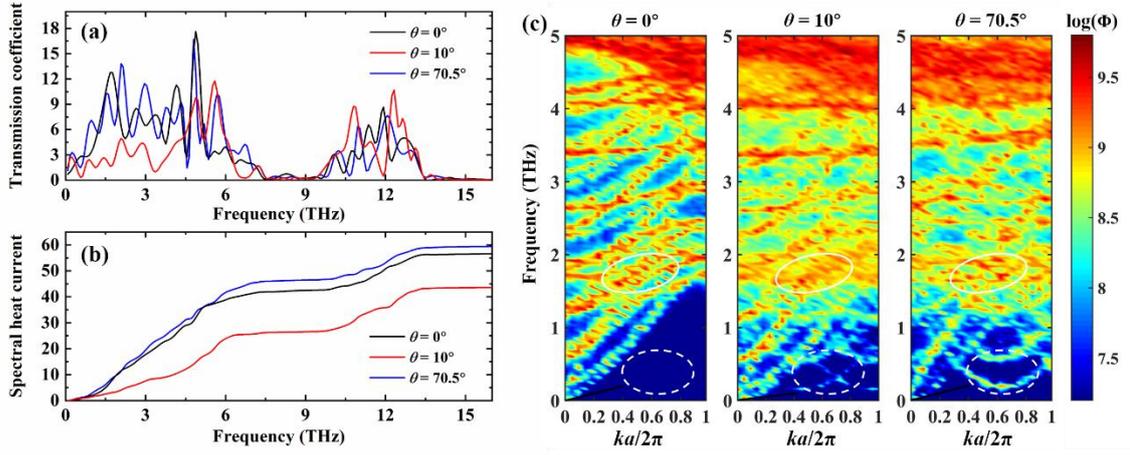

**Figure 3.** (a) Spectral transmission coefficient for TBBP with twist angle of $\theta = 0°$, $\theta = 10°$ and $\theta = 70.5°$. (b) Spectral heat current for TBBP with twist angle of $\theta = 0°$, $\theta = 10°$ and $\theta = 70.5°$. (c) SED maps for TBBP with twist angle of $\theta = 0°$, $\theta = 10°$ and $\theta = 70.5°$. The white solid ellipses denote the discrepancy of linewidth near 2 THz. The white dashed ellipses point out the appearance of new phonon brunches in non-zero twisted structures.

Out-of-plane thermal resistance

In practice, the out-of-plane thermal conduction is generally the bottleneck for heat dissipation due to the weak vdW interaction at the interface.[60, 61] Therefore, in this section, we investigated the angle dependent interfacial thermal resistance ($R$) as shown in Figure 4(a). As there is little previous study of thermal conduction in bilayer BP for reference, the calculated values of $R$ are on a reasonable order of magnitude compared to that of most 2D heterostructures (Table S6). It can be found that the interfacial thermal resistance is strongly dependent on the twist angle. When the twist angle is within 20°, the value of $R$ is low and is close to that of $\theta = 0°$. With the twist angle increasing, $R$ enhances by ~30% (at 35.3° and 64.8°) and maintains a high value from 35°-90°. Interestingly, a singularity of $R$ with a minimum value appears at 70.5°, showing a comparative value with that of $\theta = 0°$.

To elucidate the physical insights of angle-dependent interfacial thermal resistance, we first carried out the phonon density of states (PDOS) of both twisted and untwisted layers. The PDOS can be obtained by taking the fast Fourier transform of the velocity autocorrelation function (VACF) as

$$G(\omega) = \int_0^{\tau_0} \frac{\langle v(0) \cdot v(t) \rangle}{\langle v(0) \cdot v(0) \rangle} \exp(-2\pi i \omega t) dt \qquad (2)$$

where $G(\omega)$ is the normalized PDOS at the frequency $\omega$, $v(t)$ is the atomic velocity at time $t$, $\tau_0$ is the integration time, and the angle brackets denote ensemble average. The PDOS of bilayers shown in Figure 4(c) is consistent with previous calculations with spectrum frequency ranging from 0-7.5 THz and 9-13.5 THz.[62, 63] Theoretically, the phonon transport at interface is determined by the interfacial coupling strength and the matching of phonon spectra between the twisted and untwisted layers.[64] It can be clearly observed that the PDOS of two layers completely overlap when twist angle is 0°, which can be explained by that the twisted and untwisted layer is equivalent in heat transfer process. When the twist angle increases to 35.3°, the PDOS of untwisted layer remains almost unchanged while that of twisted layer exhibits a red shift in low frequency and blue shift in high frequency. The peak shift induced by the variation of vibration modes has been reported experimentally in TBG with twist-dependent Raman frequencies that are related to phonon branches of TBG.[65-67] This difference of frequency between two layers is mainly caused by the lattice mismatch and disordered vdW interaction, thus leading to the reduction in the overlap of PDOS and impeding the phonon transport across the interface. Surprisingly, the PDOS of twisted and untwisted layers overlap again when twist angle is 70.5°. Thanks to the well-ordered atomic structure of phonon magic angle, the lattice vibrations of twisted and untwisted layer are in a similar way, with the same phenomenon also reflected in AB stacking. More interestingly, this finding is in consistent with aforementioned analysis of spectrum phonon transmission, which further verifies the effective phonon transport characteristics of phonon magic angle. Therefore, the reduction in overlap of PDOS caused by the lattice mismatch and disordered atomic vibration leads to the increase of the interfacial thermal resistance. However, the recovery of crystal-like atomic vibrations in phonon magic angle re-overlaps the PDOS and thus better the interfacial transport capability.

It is well known that phonon transport properties are essentially determined by the interaction force, which is further determined by the electron cloud or the distribution of charge density. The charge density distribution is calculated to further reveal the underlying physical mechanism. Figure 4(b) shows the charge density distribution of $\theta = 0°$, $\theta = 35.3°$ and $\theta = 70.5°$. Comparing to $\theta = 35.3°$, the overlap of charge density of $\theta = 0°$ and $\theta = 70.5°$ is higher, indicating that the cohesive energy between layers is larger and so as the interfacial interaction strength (shown in Figure 4(a)). This is in agreement with the calculated interlayer distance shown in Figure S5. Both the cohesive energy and the interlayer distance exhibit the same variation tendency as the interfacial thermal resistance versus twist angle. Therefore, the overlap of charge density can be used to further explain the variation of interfacial thermal resistance, where the phonon magic angle have a highly overlapped electronic structure and strong interfacial coupling strength.

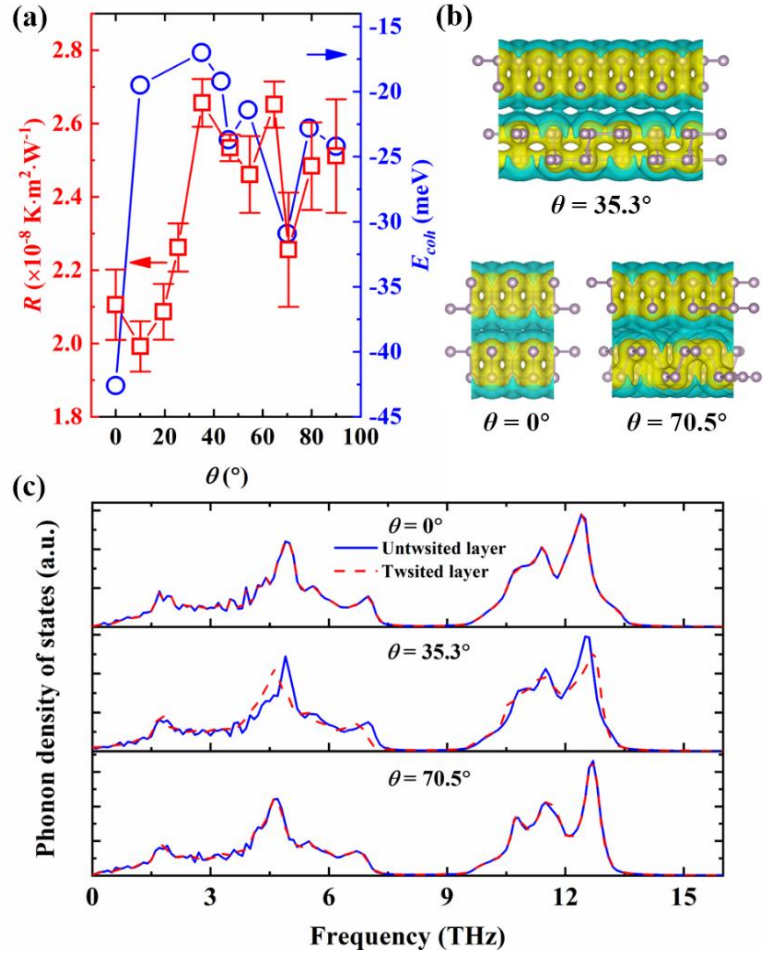

**Figure 4.** (a) Interfacial thermal resistance and cohesive energy for TBBP of various twist angles from 0° to 90°. (b) Charge density distribution of TBBP with $\theta = 0°$, $\theta = 35.3°$ and $\theta = 70.5°$. (c) Phonon density of states of untwsited layer and twisted layer with $\theta = 0°$, $\theta = 35.3°$ and $\theta = 70.5°$, in which yellow and blue refer to charge density contour with the isosurface level of 0.05 and 0.006, respectively. Pink represents the P atom.

CONCLUSION

In conclusion, we have found a 'phonon magic angle' in twisted 2D puckered materials by molecular dynamics simulation. Comparing to the declaration of phonon transport capability for most twist angles, the magic angle remains a high level or even enhanced due to van der Waals confinement betweem layers. This novel phenonemon is originated from the ordered vdW interaction and phonon vibration caused by the perfect lattice arrangement that the atoms of the top layer can get stuck to the spaces of the bottom layer. As a typical example, TBBP is taken to systematically investigate the in-plane and out-of-plane phonon transport properties of twisted 2D puckered materials. The spectral phonon transmission coefficient and sepectral energy density reveal that the suppression of low frequency acoustic phonons and the reduction of phonon lifetimes and group velocity lead to the reduction of in-plane thermal conductivity. Moreover, the decrease of interfacial thermal conductance is resulted from the reduction in the overlap of the PDOS and charge density. These reductions can be essentially attributed to the lattice mismatch and disordered phonon vibration because of interlayer twsit. As a striking contrast, the phonon magic angle serves as a singularity due to its ordered atomic structure similar to AB stacking. The finding of 'phonon magic angle' in 2D puckered materials offers a new insight to manipulate the vdW interaction and paves the way to the phononic, electronic and photonic modulation in low-dimensional asymmetrical materials.

METHODS

We use the Large-scale Atomic/Molecular Massively Parallel Simulator (LAMMPS) software package[68] to calculate the thermal transport properties of TBBP. The relative twist angle between layers is constructed by means of coincidence site lattice (CSL) theory[69] which ensures the misoriented bilayer atomic structure remains periodic. Considering the in-plane anisotropy of BP, we fix the lattice orientation of one layer to be consistent with the coordinate axes and twist the other layer to a certain angle, otherwise the direction of heat transfer in TBBP cannot be defined. As shown in Figure 2(a), the zigzag and armchair direction of top layer is correspond to the *x*-axis and *y*-axis, respectively, in all twist cases. Twelve twist angles are chosen ranging from 0° in AB stacking to 90° when both crystal axes of two layers are orthogonal by applying twist to the bottom layer (shown in Figure S7). The angles varying from 90° to 180° are not taken into consideration because they are symmetrical with the former.

In the simulation, the Stillinger-Weber (SW) potential parameterized by Jiang et al.[70] is employed to describe the covalent interaction between P-P atoms. This SW potential can successfully reproduce the phonon dispersions that agree well with the first-principle results and is widely used in MD simulation of BP.[63, 71, 72] The interlayer long-range interaction, namely van der Waals (vdW) interaction, is modeled using 12-6 Lennard-Jones (LJ) potential, as

$$V(r) = 4\varepsilon_{P-P} \left[ \left( \frac{\sigma_{P-P}}{r_{ij}} \right)^{12} - \left( \frac{\sigma_{P-P}}{r_{ij}} \right)^{6} \right] \quad (3)$$

where $\varepsilon_{P-P}$ is the depth of the potential well, $\sigma_{P-P}$ is the zero potential distance and $r_{ij}$ is the separation distance between atoms $i$ and $j$. The standard parameters adopted in present simulation is based on the Universal Force Field (UFF) model[73] with $\varepsilon_{P-P}$ = 0.0132 eV and $\sigma_{P-P}$ = 3.695 Å while the cutoff distance of the L-J potential is set to 15 Å, consistent with the literature.[72]

Figure S8(a) represents the diagram of non-equilibrium molecular dynamics (NEMD) simulation to calculate the in-plane thermal conductivity. Periodic boundary condition is applied in width direction to eliminate the edge effect and atoms at two ends of length direction is fixed in space. A 10 nm vacuum layer is added in $z$ direction to avoid the atomic interaction with the system image. Langevin heat baths[74, 75] are applied at atoms adjacent to the fixed ends to form a temperature gradient with high temperature maintained at $T_h$ = 330 K and low temperature at $T_c$ = 270 K. We apply the heat baths to $x$ and $y$ directions, respectively, to extract the thermal conductivity of both directions. The system length is set as 20 nm in length and 6 nm in width based on the fact that the thermal conductivity convergences at 5 nm with the increasing of width (shown in Figure S9).

The velocity-Verlet algorithm[76] is used to integrate the differential equations of motions with a time step of 0.5 fs for all cases. Before NEMD simulations, the initial system is relaxed in the isothermal-isobaric (NPT) ensemble for 0.5 ns to release the internal stress. Then the system is placed under Nosé-Hoover thermostat to reach equilibrium at the designated temperature of 300 K. After relaxing in the canonical ensemble (NVT) for 0.5 ns, the system is then switched to the microcanonical ensemble (NVE) for another 0.5 ns. Following equilibration, the thermal baths are conducted to perform the NEMD simulation for long enough time of 6 ns to ensure that the steady state is achieved. In-plane thermal conductivity is calculated according to the Fourier law, as

$$\kappa = -\frac{J/A}{dT/dx} \quad (4)$$

where $A$ is the cross-sectional area, defined as the product of width and thickness. The thickness of monolayer BP is chosen as 5.25 Å which has been used in previous calculations.[63, 70, 71] $J$ and $dT/dx$ are the heat flux transported in the system and temperature gradient, respectively. The heat flux ($J$) is recorded by the average of the input/output power at the two baths and calculated as

$$J = \frac{\Delta E_{in} + \Delta E_{out}}{2\Delta t} \quad (5)$$

where $\Delta E_{in}$ and $\Delta E_{out}$ are the energy added and removed from each bath at each time step $\Delta t$. We record the input/output power and temperature profile in untwisted layer, twisted layer and bilayer to calculate their thermal conductivity, respectively. Both the heat flux and temperature gradient are extracted by the average value over the last 1 ns simulations when the temperature gradient is well established and the heat current is independent of simulation time, as shown in Figure S8(b) and (c).

To explore the out-of-plane thermal transport capability, we extract the interfacial thermal resistance by transient heating technique, mimicking the experimental pump-probe method. This method has been successfully employed to study the interfacial thermal resistance of various 2D heterostructures in MD simulations.[62, 77-79] The system is firstly relaxed under NPT, NVT and NVE ensemble for 0.5 ns, respectively. After fully equilibrated at the specified temperature, an ultrafast heat pulse is added to the untwisted layer for 50 fs to raise the temperature to 500 K while the temperature of the twisted layer remains unchanged. Then the system is relaxed and reaches a new steady state with heat dissipating from hot source to cold reservoir across the interface. The total energy of each layer is recorded every time step for 50 ps. During this process, the whole system is in vacuum, so the exchange heat flux through interface is correlated to the energy variations in untwisted layer, which can be expressed as

$$\frac{\partial E_t}{\partial t} = \frac{A(T_{untwist} - T_{twist})}{R} \tag{6}$$

where $E_t$ is the total energy of untwisted layer at time $t$, $A$ is the in-plane area, $T_{untwisted}$ and $T_{twisted}$ are the temperature of untwisted and twisted layer, respectively. The evolutions of temperature and total energy after 50 fs heat impulse are shown in Figure S10(a). By integral with respect to time *t*, the Equation (6) can be transferred as

$$E_t = E_0 + \frac{A}{R}\int_0^t (T_{untwist} - T_{twist})dt \qquad (7)$$

where $E_0$ is the initial energy of the untwisted layer. Figure S10(b) exhibits the correlation between $E_t$ and integral term. It is clear that the thermal resistance, *R*, can be easily obtained by the slope (*A/R*) by a linear fitting. To suppress the noise, the data collected is averaged over every 50 time steps. The system size is set to 10 nm×10 nm to ensure the size independent thermal resistance (shown in Figure S11). All the results in this work is ensemble averaged over six independent calculations with different initial conditions.

Supporting Infromation

Thermal conductivity of TBBP along x direction, phonon dispersion relation and group velocity for monolayer BP, calculation for spectral phonon transmission coefficient, calculation for spectral energy density, interlayer distance for TBBP of various twist angles, a summary of various interfacial thermal resistance, lattice structure of TBBP with twelve twist angles, schematic diagram of NEMD simulation configuration, verification of width independent thermal conductivity, temperature and energy evolution of transient heating method, verification of size independent interfacial thermal resistance.

AUTHOR INFORMATION

**Corresponding Author**

*E-mail: x-zhang@mail.tsinghua.edu.cn

**Notes**

The authors declare no competing financial interest.

ACKNOWLEDGMENT

This work was supported by the National Natural Science Foundation of China (Grant Nos. 51636002 and 51827807), China Postdoctoral Science Foundation 2020M670321 and Natural Science Foundation of Shaanxi Province (2020JQ-629)

Table of contents

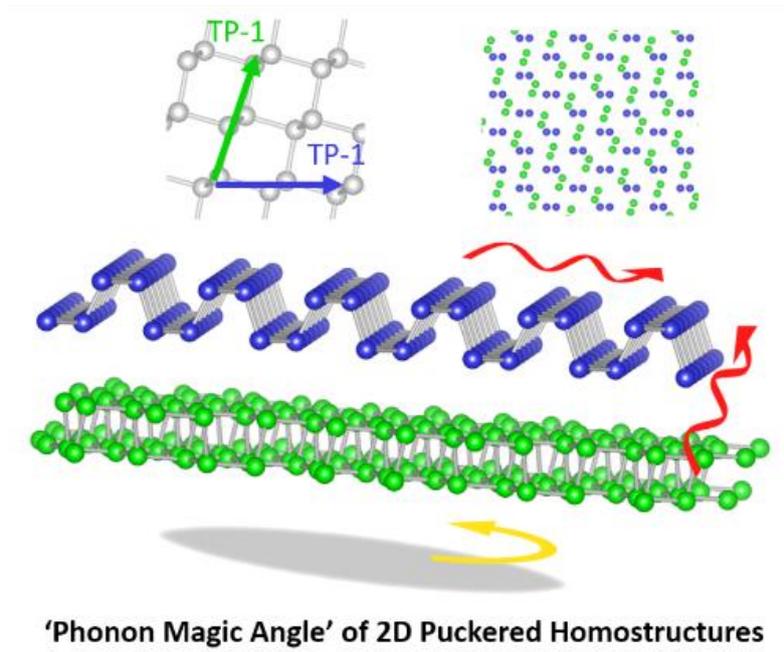

'Phonon Magic Angle' of 2D Puckered Homostructures

Supporting Information

# Phonon Magic Angle in Two-Dimensional Puckered Homostructures

*Yufeng Zhang[†, §], Meng An[†, ‡, §], Dongxing Song[†], Haidong Wang[†], Weigang Ma[†],*

*Xing Zhang[*, †]*

[†] Key Laboratory for Thermal Science and Power Engineering of Ministry of

Education, Department of Engineering Mechanics, Tsinghua University, Beijing

100084, China

[‡] College of Mechanical and Electrical Engineering, Shaanxi University of Science

and Technology, Xi'an 710021, China

[§] Y. Z. and M. A. contributed equally to this work

[*] Corresponding author. E-mail: x-zhang@mail.tsinghua.edu.cn

List of supporting information:

**Part S1.** Thermal conductivity of TBBP along *x* direction.

**Part S2.** Phonon dispersion relation and group velocity for monolayer BP.

**Part S3.** Calculation for spectral phonon transmission coefficient

**Part S4.** Calculation for spectral energy density

**Part S5.** Interlayer distance for TBBP of various twist angles.

**Part S6.** A summary of various interfacial thermal resistance.

**Part S7.** Lattice structure of TBBP with twelve twist angles.

**Part S8.** Schematic diagram of NEMD simulation configuration.

**Part S9.** Verification of width independent thermal conductivity.

**Part S10.** Temperature and energy evolution of transient heating method.

**Part S11.** Verification of size independent interfacial thermal resistance.

**Part S12.** Auxiliary animation materials for thermodynamic process.

**Part S1.** Thermal Conductivity of TBBP Along *x* Direction

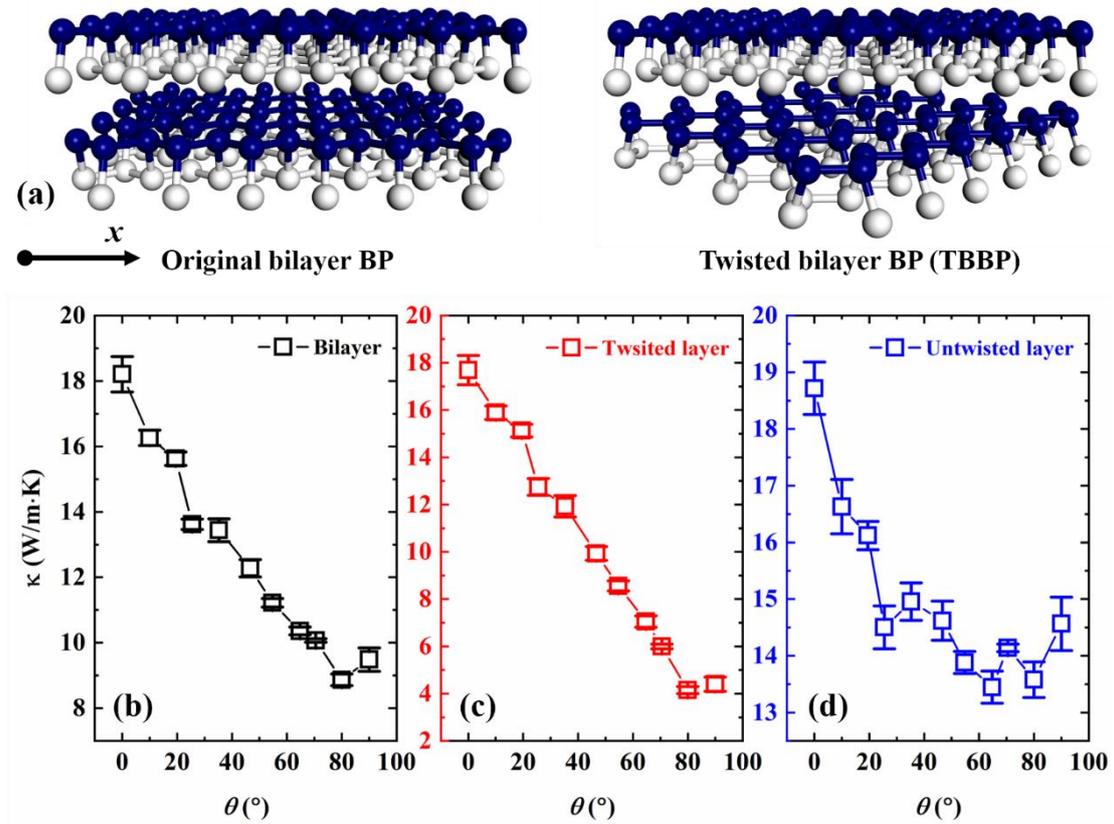

**Figure S1.** (a) Concept diagram for Original bilayer BP and TBBP. The top layer (untwisted layer) is always fixed with zigzag and armchair direction corresponding to the *x*-axis and *y*-axis, respectively The different colors of P atoms are for clearly distinguishing the morphology. Below are the thermal conductivity results for (b) bilayer, (c) twisted layer and (d) untwisted layer of various twist angles from 0° to 90° along the *x* direction, respectively.

**Part S2.** Phonon dispersion relation and group velocity for monolayer BP

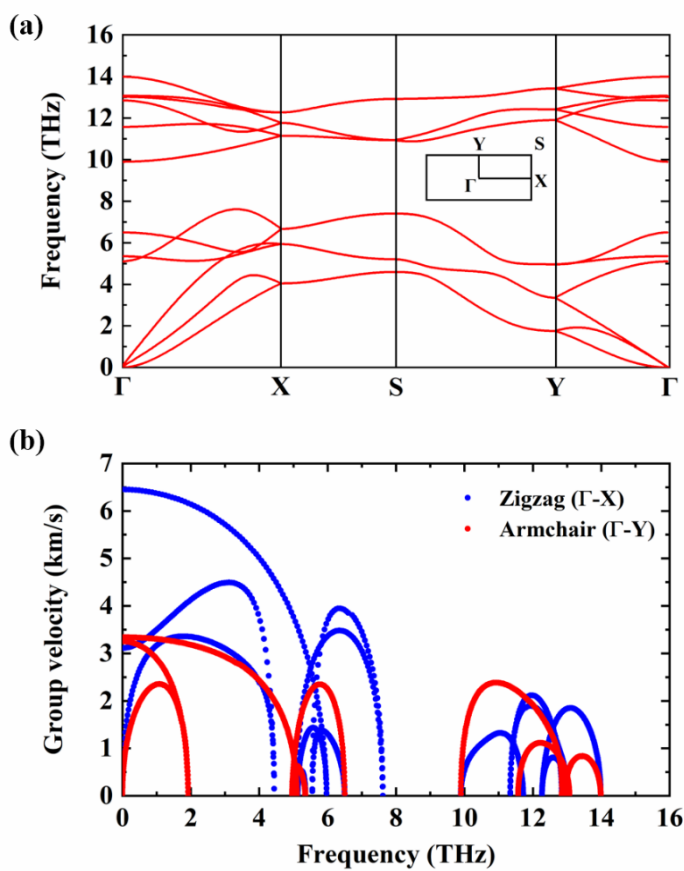

**Figure S2.** (a) Phonon dispersion along the path passing through the high-symmetry k-point in the Brillouin zone (BZ) of monolayer BP. The Γ-X and Γ-Y directions are corresponding to the Zigzag and armchair directions, respectively. (b) Phonon group velocity of Zigzag (blue) and armchair (red) directions. The above vibrational eigenmode analysis of phonon in monolayer BP is implemented by General Utility Lattice Program (GULP).[S1]

**Part S3.** Calculation for Spectral Phonon Transmission Coefficient

The calculation of spectral phonon transmission coefficient $T(\omega)$ is according to the equation defined by Sääskilahti et al.[S2, S3] as

$$T(\omega) = \frac{q(\omega)}{k_B \Delta T} \quad (S1)$$

where $k_B$ is the Boltzmann parameter and $\Delta T$ is the temperature difference between the two Langevin thermostats. $q(\omega)$ is the frequency dependent spectral heat flux across the imaginary cross-section (chosen in the middle of the heat flow direction here), which can be derived as

$$q(\omega) = \frac{2}{St_s} \mathrm{Re} \sum_{i \in L} \sum_{i \in R} \langle \mathbf{F}_{ij}(\omega) \cdot \mathbf{v}_i(\omega)^* \rangle \quad (S2)$$

where $t_s$ is the simulation time, and $\mathbf{F}_{ij}$ is the inter-atomic force on atom $i$ due to atom $j$. Here, "$L$" and "$R$" denotes the left and right segment, respectively, which are located at two sides of the imaginary cross-section.

**Part S4.** Calculation for Spectral Energy Density

The expression of the spectral energy density (SED) analysis $\Phi(\mathbf{k}, \omega)$ in our calculation is defined as[S4, S5]

$$\Phi(\mathbf{k},\omega) = \frac{m}{4\pi n \tau_0} \sum_{\gamma} \sum_{b=1}^{B} \left| \int_0^{\tau_0} \sum_{l=1}^{n} u_\gamma(l,b,t) \times \exp(i\mathbf{k} \cdot \mathbf{r}_l - 2\pi i v t) dt \right|^2 \quad (S3)$$

where $\mathbf{k}$ is the wave vector, $\omega$ is frequency, $m$ is the mass of carbon atom, $n$ is the number of unit cells, $r_l$ is the equilibrium position of the each unit cell, $b$ is the atom index in the each unit cell, $u_\gamma$ is the atom velocity in the $\gamma$ direction, and $\tau_0$ is the total simulation time. The atomic velocities were obtained from NVE (constant mass, volume, and energy) ensemble at a temperature of 300 K. The covalent and non-bonded interactions are described by SW and L-J potential, respectively, and periodic boundary conditions are imposed in all directions, which are in agreement with the settings employed to calculate the thermal conductivity. We integrate the equations of motion using the velocity Verlet scheme with a 0.5 fs time step and collect the data every 10 steps. The SED in the full-frequency range is shown in Fig. S9.

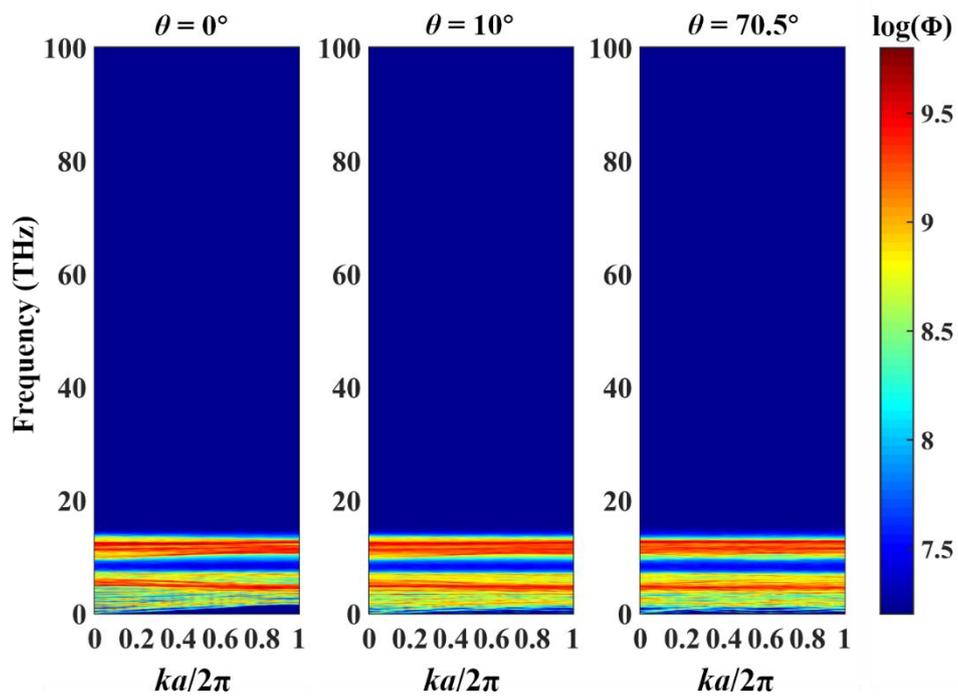

**Figure S4.** The SED maps for $\theta = 0°$, $\theta = 10°$ and $\theta = 70.5°$ in the full-frequency range

**Part S5.** Interlayer Distance for TBBP of Various Twist Angles

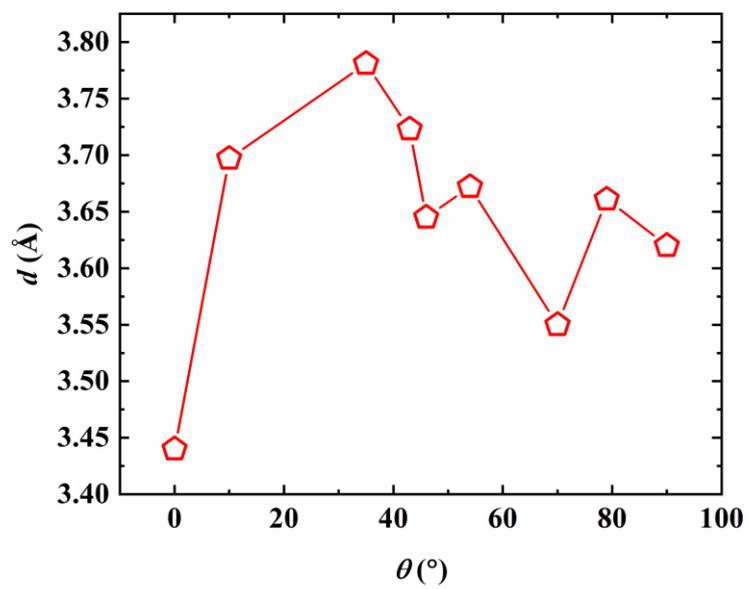

**Figure S5.** Interlayer distance for TBBP of various twist angles from 0° to 90°.

**Part S6.** A Summary of Various Interfacial Thermal Resistance

**Table S6.** A summary of interfacial thermal resistance of several popular 2D heterostructures or homostructures from MD simulation

| References | Materials | $T$ (K) | $R$ (×10$^{-8}$ K·m$^2$·W$^{-1}$) |
|---|---|---|---|
| Liu et al.[S6] | Graphene/Silicene | 200-700 | ~4.2-16.8 |
| Wei et al.[S7] | Graphene/Graphene | 400-1200 | ~0.021-0.025 |
| Zhang et al.[S8] | Graphene/h-BN | 200-700 | 9.04-29.6 |
| Liu et al.[S9] | Graphene/MoS$_2$ | 200-500 | ~8.3-25 |
| Hong et al.[S10] | Graphene/phosphorene | 50-350 | 7.55-17.34 |
| Wang et al.[S11] | C$_2$N/C$_2$N | 300 | 3.4 |
| Zhang et al.[S12] | MoS$_2$/MoSe$_2$ | 300 | 33.9-35.6 |
| **Our work** | **BP/BP** | **300** | **1.99-2.66** |

**Part S7.** Lattice Structure of TBBP with Twelve Twist Angles

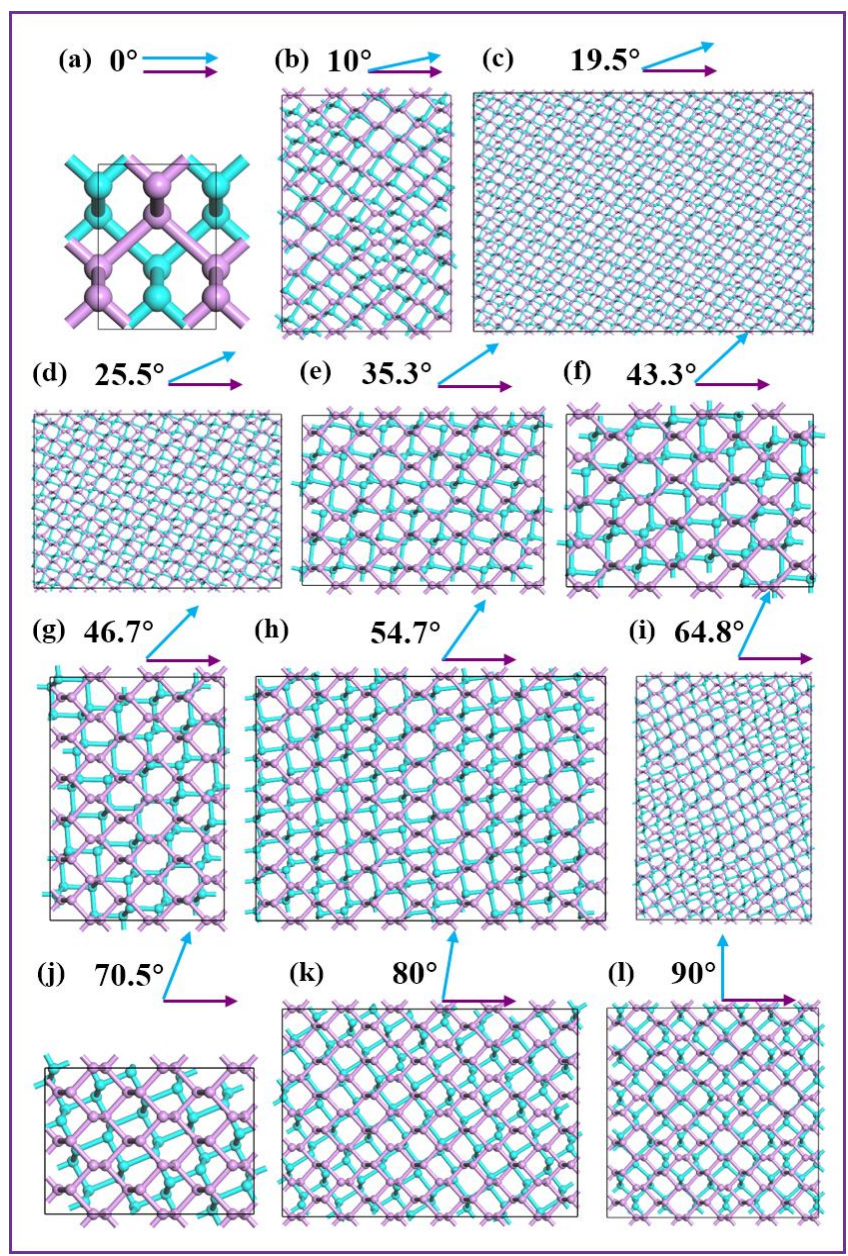

**Figure S7.** Lattice structure of TBBP with twist angle of (a) $\theta = 0°$, (b) $\theta = 10°$, (c) $\theta = 19.5°$, (d) $\theta = 25.5°$, (e) $\theta = 35.3°$, (f) $\theta = 43.3°$, (g) $\theta = 46.7°$, (h) $\theta = 54.7°$, (i) $\theta = 64.8°$, (j) $\theta = 70.5°$, (k) $\theta = 80°$ and (l) $\theta = 90°$. The pink and blue arrows mark the armchair direction of untwisted and twisted layer, respectively.

**Part S8.** Schematic Diagram of NEMD Simulation Configuration

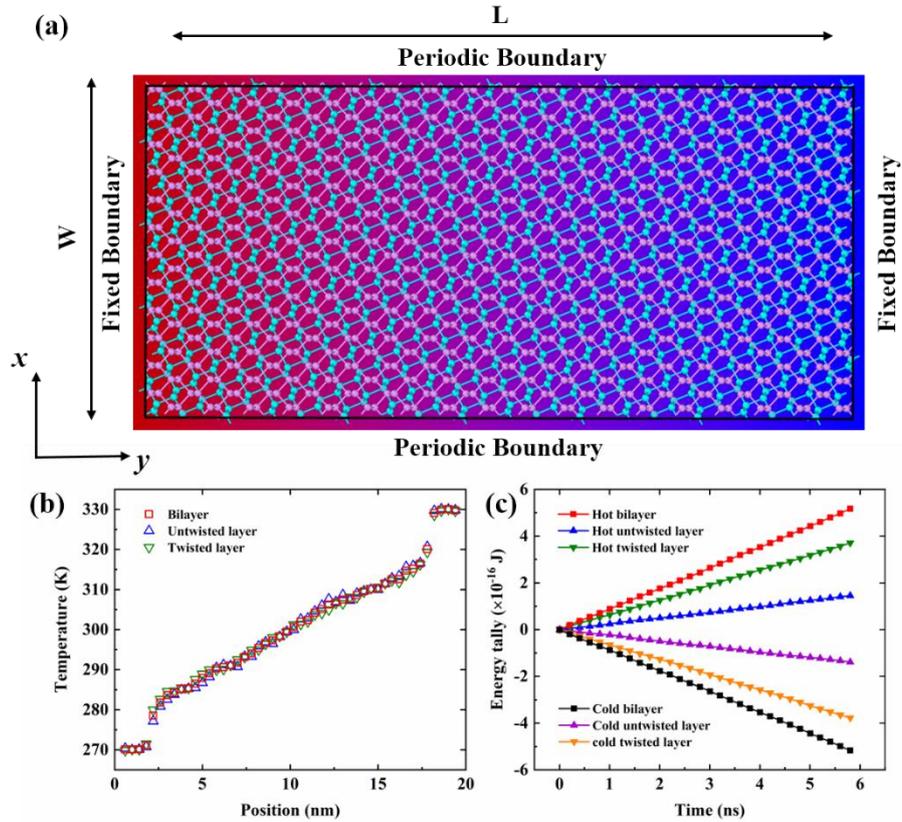

**Figure S8.** (a) The schematic diagram of NEMD simulation configuration for calculating the thermal conductivity of TBBP with $\theta = 70.5°$ along the $y$ direction. The fixed (periodic) boundary condition is applied along the $y$ ($x$) direction. The temperature of heat source (heat sink) is set to $T_h = 330$ K ($T_c = 270$ K). (b) The corresponding temperature profile of the untwisted layer, twisted layer and bilayer, respectively. The identical temperature profiles indicate the negligible heat leakage between layers. (c) The energy tallies through cross-section area of untwisted layer, twisted layer and bilayer, respectively, during 6 ns simulations. The heat current is taken as the slope of the curves.

**Part S9.** Verification of Width Independent Thermal Conductivity

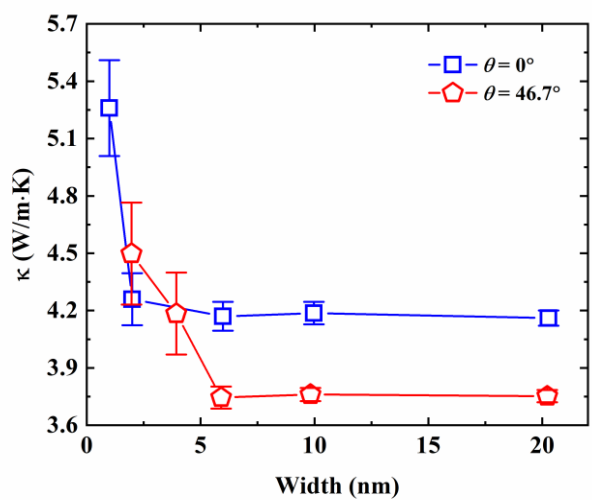

**Figure S9.** The thermal conductivity of untwisted layer along the *y* direction with respect to various width for TBBP ($\theta$ = 46.7°) and BP and original bilayer BP ($\theta$ = 0°), respectively. The twist angle of 46.7° is selected to illustrate the difference more clearly.

**Part S10.** Temperature and Energy Evolution of Transient Heating Method

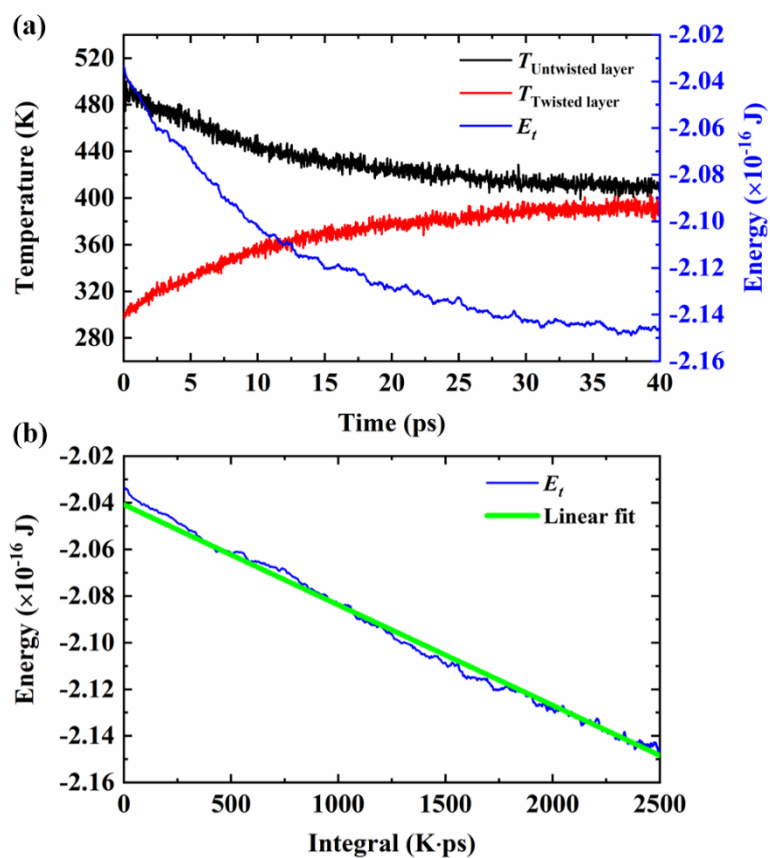

**Figure S10.** (a) Temperature evolutions of untwisted layer and twisted layer, respectively, and energy profile of untwisted layer after 50 fs heat impulse. (b) By integrating the temperature differences between $T_{\text{Untwisted layer}}$ and $T_{\text{Twisted layer}}$, the energy relaxation profile of untwisted layer can be correlated to integral term ($\Delta T \cdot \mathrm{d}t$) directly. The slope can be linearly fitted (green line) to extract the interfacial thermal resistance.

**Part S11.** Verification of Size Independent Interfacial Thermal Resistance

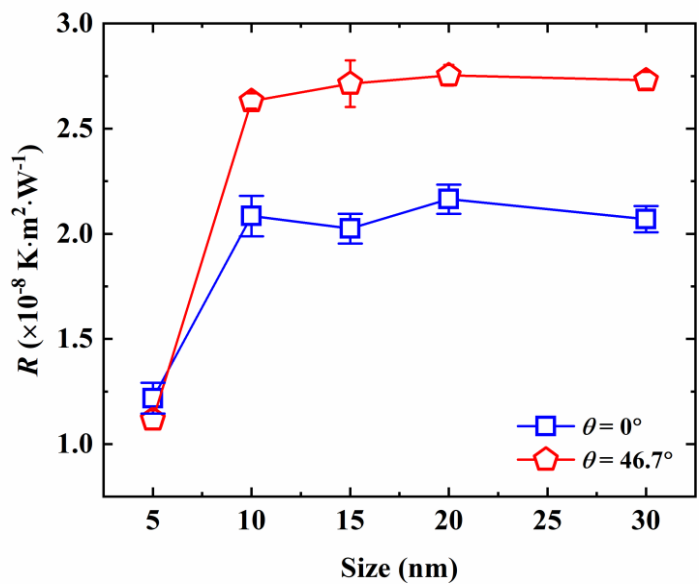

**Figure S11.** The interfacial thermal resistance of with respect to various size for TBBP ($\theta$ = 46.7°) and original bilayer BP ($\theta$ = 0°), respectively. The twist angle of 46.7° is selected to illustrate the difference more clearly.

**Part S12.** Auxiliary Animation Materials for Thermodynamic Process

The following two videotapes visually illustrate the drifting-induced thermodynamic process of stabilized stacking of AB and magic angle with the same initial configuration.

**AB-BP.avi** recorded the comprehensive thermodynamics process for BP to reach the stabilized AB stacking.

**Magic-BP.avi** recorded the comprehensive thermodynamics process for BP to reach the stabilized stacking of magic angle.